\documentclass[aps,pra,twocolumn,showpacs,groupedaddress,amsmath,amssymb]{revtex4}

\usepackage{graphicx}
\graphicspath{{b:/officestuff/papers/nonlin_LRC/figures/}}

\usepackage{color}
\usepackage{comment}
\usepackage{bm}
\usepackage{latexsym}

\begin{document}

\title{Optical Asymmetry Induced by ${\cal PT-}$symmetric Nonlinear Fano Resonances}
\author{F. Nazari$^{1,2}$,N. Bender$^1$, H. Ramezani$^1$, M. K. Moravvej-Farshi$^{2}$, D. N. Christodoulides$^3$, T. Kottos$^1$} 
\affiliation{$^1$Department of Physics, Wesleyan University, Middletown, CT-06459, USA}
\affiliation{$^2$Faculty of Electrical \& Computer Engineering, Tarbiat Modares University, Tehran 1411713116, Iran}
\affiliation{$^3$College of Optics \& Photonics-CREOL, University of Central Florida, Orlando, Florida 32816, USA}
\date{\today}

\begin{abstract}
We  introduce a new type of Fano resonances, realized in a photonic circuit which consists of two nonlinear ${\cal PT}$-symmetric micro-resonators 
side-coupled to a waveguide, which have line-shape and resonance position that depends on the direction of the incident light. We utilize these features 
in order to induce asymmetric transport up to $47$ dBs in the optical C-window. Our set-up requires low input power and does not compromise 
the power and frequency characteristics of the output signal.
\end{abstract}

\maketitle


The realization of micron scale photonic elements and their integration into a single chip-scale device constitute an important challenge,
both from a fundamental and a technological perspective \cite{ST91}. An important bottleneck towards their realization is achieving on-
chip optical isolation, that is the control of light propagation in predetermined spatial directions. Standard approaches for optical isolation 
rely mainly on magneto-optical (Faraday) effects, where space-time symmetry is broken via external magnetic fields. This 
approach requires materials with high Verdet constants and/or large size non-reciprocal structures which are incompatible with on-chip 
integration \cite{ST91}. Alternative proposals, for the realization of optical diodes, 
include dynamical modulation of the index of refraction \cite{YF09}, the use of opto-acoustic effects \cite{KBR11}, and optical non-linearities 
\cite{GAPF01,SDBB94,LC11,FWVSNXWQ12,BFBRCEK13}. Most of these schemes, have serious drawbacks which make them unsuitable 
for small-scale implementation. In some of these cases, complicated designs that provide structural asymmetry are necessary, or the 
transmitted signal has different characteristics (e.g different frequency) than the incident one. In other cases, direct reflection or absorption  
dramatically affects the functionality leading to an inadequate balance between transmitted optical intensities and figures of merit. 

Recently, optical microresonator structures \cite{V03} with high-quality factors that trigger non-linear effects, have attracted increasing 
attention as basic elements for the realization of on-chip optical diodes \cite{FWVSNXWQ12,BFBRCEK13}. The basic geometries used 
consists of two waveguides coupled with two single mode non-linear cavities. These geometries typically allow for a narrow band 
transmission channel with a symmetric Lorentzian transmittance lineshape. In Ref.~\cite{FWVSNXWQ12} the structure was passive and 
the diode action was imposed due to the asymmetric coupling of the cavities to the waveguides. The drawback of this protocol is that 
high degree of asymmetric transport (due to strong asymmetric coupling) is achieved at the expense of low intensity output signals. 
On the other hand, the proposal of Ref.~\cite{BFBRCEK13} involved active cavities (one with gain and another with loss) which  
excite the nonlinear resonances differently, depending on the incident direction. This latter proposal has been demonstrated recently in 
a beautiful experiment \cite{POLMGLFBY13} where gain in the first resonator is supplied by optically pumping Erbium ions embedded in 
a silica matrix while the second resonator exhibits passive loss. However, the degree of transport asymmetry is moderate and 
it is achieved only for high values of gain.

\begin{figure}
\includegraphics[width=1\linewidth, angle=0]{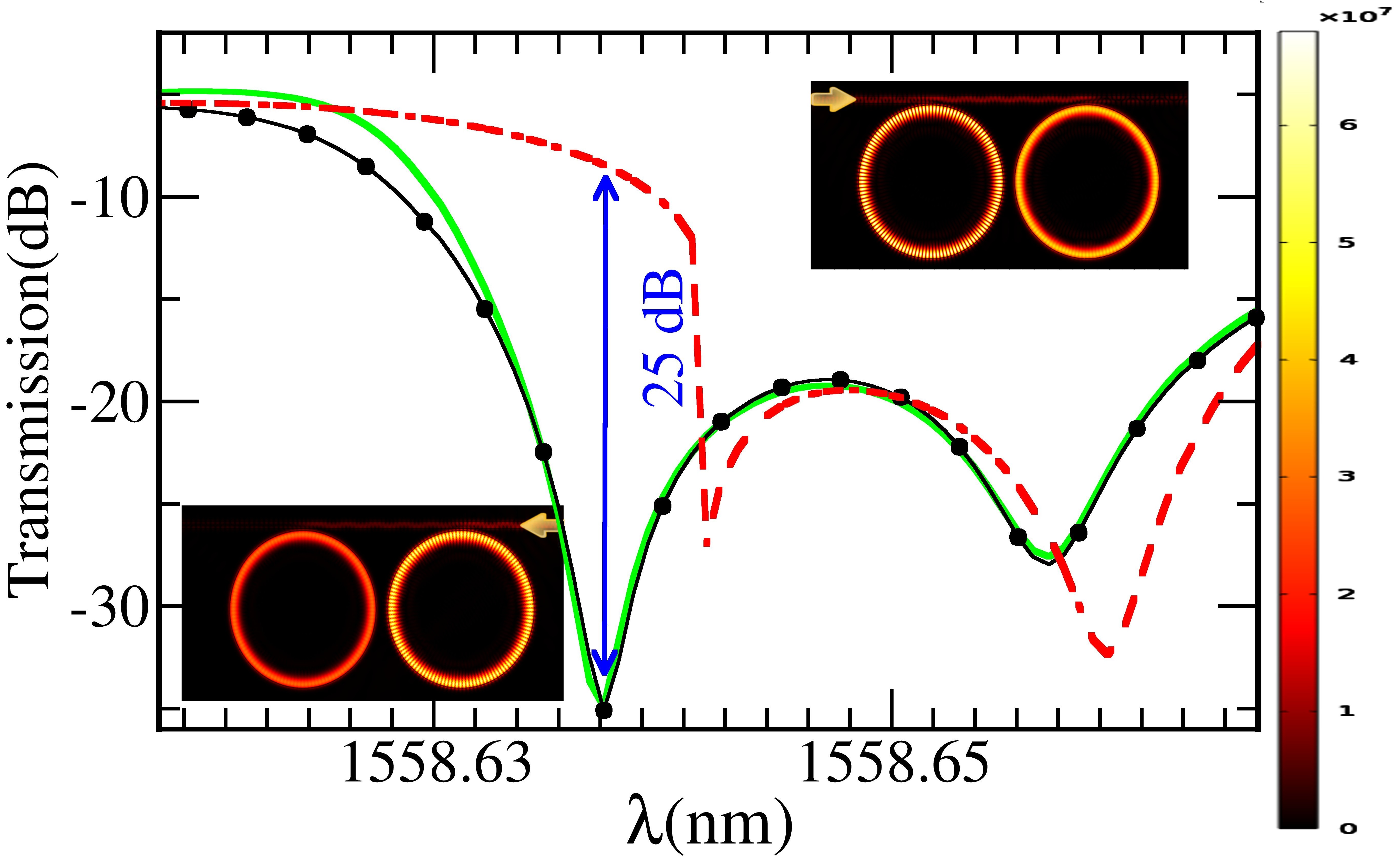}
\caption{Main panel: The transmittances $T(\lambda)$ of the linear system for incident waves from the gain (filled black circles) and 
from the loss (black line) side are compared with the corresponding transmittances in the case of microdiscs with Kerr non-linearity. 
In the latter case, the resonance from the loss side (green line) has experienced a small red-shift with respect to the linear structure. 
In contrast, the transmittance curve (both line-shape and resonance position) of an incident wave entering the structure from the gain 
side (red dotted-dashed line) is different. Insets: A schematic of the photonic circuit that generates directional ${\cal PT}$-symmetric 
Fano resonances.}
\label{fig:fig1}
\end{figure}

In this Letter we investigate the possibility to utilize a new type of nonlinear Fano resonances emerging in a parity-time (${\cal PT}$) symmetric 
framework in order to create asymmetric transport. The proposed photonic 
circuit consists of two non-linear ${\cal PT-}$symmetric microcavities which are side-coupled to a single waveguide. We show that this system 
naturally exhibits Fano resonances \cite{F61} which, due to the interplay of non-linearity with the active elements, are triggered at different 
resonance frequencies and have different lineshape, depending on the direction of the incident light. Fano resonances, sometimes behaving like 
{\it coupled-resonator-induced transparency} \cite{SCFRB04}, were first introduced in the optics framework  in Refs. \cite{HL91,XLLY00}. Their shape is distinctly 
asymmetric, and differs from the conventional symmetric Lorentzian resonance curves (for a recent review see \cite{MFK10}). This asymmetric 
resonance profile essentially  results from the interference between a direct and a resonance-assisted indirect pathway \cite{SCFRB04}.

A realization of the proposed photonic diode is shown in the inset of Fig. \ref{fig:fig1}. For demonstration purposes, let the core of both the 
waveguide and of the two microdiscs to be of AlGaAs material. The permittivity of both microdisk resonators and of the waveguide is taken 
to be $\epsilon'=11.56$ while the nonlinear Kerr coefficient for the microdiscs is $\chi=1e-19 (m^2 /V^2)$. The radius of the mircrodisks and 
their distance of each other are $5 \mu m$ and $770 nm$, respectively. Moreover, the width of the waveguide and its coupling distance 
to the resonators are $460 nm$ and $120 nm$, respectively. The circuit is operated at the optical communication window at wavelength 
around $\lambda\approx 1558.6 nm$ with one disc experiencing gain, while the other one having an equal amount of loss described 
by the imaginary part of the permitivity $\epsilon''=0.00063$. The structure is invariant under ${\cal PT}$ symmetry where the ${\cal P}$ is
the parity reflection, with respect to the axis of symmetry located at the middle between the two resonators, and ${\cal T}$ is the time reversal
operator which turns loss to gain and vice versa. The concept of ${\cal PT}-$ symmetry first emerged within the context of 
mathematical physics. In this regard, it was recognized that a class of non-Hermitian Hamiltonians that commute with the ${\cal PT}$ 
operator may have entirely real spectra \cite{BB98}. Lately, these notions have been successfully migrate and observed 
in other areas like photonic \cite{MGCM08,RMGCSK10,GSDMVASC09,ZCFK10,L10b,LRKCC11,RKGC10} and electronic circuitry \cite{BFBRCEK13,SLZEK11,LSEK12}. 

In Fig.\ref{fig:fig1} we show some transport simulations using COMSOL modeling. The input power used in the simulations is $P=1.2 mW$. 
We find that the left-to-right transmittance $T_L(\lambda)$ differs from the right-to-left transmittance $T_R(\lambda)$, i.e. 
$T_L\neq T_R$. The asymmetric transport is most pronounced near the Fano resonances $\lambda_{\chi=0}^{\cal PT}$ 
of the linear structure, and constitutes our main result. We stress that non-reciprocal transport is strictly forbidden by the Lorentz reciprocity 
theorem in the case of linear, time-reversal symmetric systems \cite{note1}.   At the same time, it cannot be achieved neither by a conservative 
nonlinear medium by itself nor by linear ${\cal PT}$-symmetric structures (see black filled circles in Fig. \ref{fig:fig1}) \cite{LSEK12}.

\begin{figure}
\includegraphics[width=1\linewidth, angle=0]{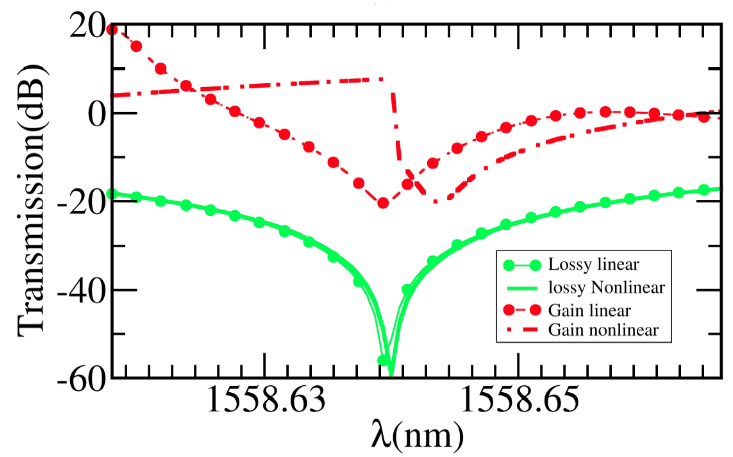}
\caption{Transmittance curves $T(\lambda)$ for a single gain or loss microdisc side-coupled to a waveguide. The filled red circles 
(dashed-dotted line) correspond to a gain disc in the absence (presence) of  Kerr nonlinearity. The filled green circles (green bold line) 
correspond to a lossy disc in the absence (presence) of Kerr nonlinearity. In the case of a gain disc a red-shift of the resonance 
position and a strong modification of the transmittance line-shape is observed.
}
\label{fig:fig2}
\end{figure}

Next we analyze the origin of the asymmetry between left and right transmittances, near the Fano resonances $\lambda_{\chi=0}^{\cal PT}$ 
of the linear photonic circuit of Fig. \ref{fig:fig1}. To understand better its origin, we first discuss the transport characteristics of a single gain 
(lossy) non-linear microdisc side-coupled to a waveguide. In the case that the incident light traveling the waveguide couples with a gain 
resonator it will be amplified substantially because of the interaction with the gain medium, and the high $Q$ factor of the disc. 
Consequently, the signal has sufficiently high power to trigger the non-linearity and red-shift the disc's resonance $\lambda_{\chi}^{G}$, thus 
allowing it to pass with small (or even not at all) attenuation at the resonance wavelength $\lambda_{\chi=0}^{G}$ of the gain cavity (dashed 
line in the inset of Fig \ref{fig:fig2}). On the other hand, when light couples to a lossy microdisc, the optical energy stored in this disc is not high 
enough to appreciably red-shift (via non-linearity) the resonance because of the power reduction due to the losses. As a result the transmittance 
has a resonance dip at $\lambda\approx \lambda_{\chi=0}^{L}$. Obviously in both cases discuss here, we have symmetric transport i.e. 
$T_L(\lambda)=T_R(\lambda)$.

When both linear microdiscs, i.e. the one with gain (left) and the other one with loss (right), are side-coupled to the waveguide the  transmittance
shows a peak in the middle of the resonant dip. This phenomenon is an optical analogue \cite{XSPSFL06} of electromagnetically induced 
transparency (EIT) and it is known as {\it coupled-resonator-induced transparency} \cite{SCFRB04}. It is associated with the 
interaction between two Fano resonances with spectral widths which are comparable 
to or larger than the frequency separation between them. These Fano resonances have been formed due to coherent interferences between the 
two coupled resonators. Still we observe that left and right transmittance are equal i.e. $T_L(\lambda)=T_R(\lambda)$.

When nonlinearities are considered, the transmission near the Fano resonances is asymmetric and depends strongly on the direction of the 
incident light. Below we concentrate on the wavelength domain on the left of the transparent window where, for our set up, the asymmetry is stronger. 
In this case the incident light entering the waveguide from the left is first coupled to the gain resonator which amplify the light intensity; 
thus inducing optical nonlinearity of the material. As a result, the resonance wavelength $\lambda_{\chi}^{\cal PT}$ of the non-linear 
${\cal PT}$-structure is strongly red-shifted with respect to the resonance wavelength $\lambda_{\chi=0}^{\cal PT}$ of the linear ${\cal PT}$ 
circuit. Thus at incident wavelength $\lambda=\lambda_{\chi=0}^{\cal PT}$ the photonic circuit of Fig. \ref{fig:fig1} is almost transparent. 
Moreover, the outgoing signal is strong due to the amplification at the gain disc and despite the fact that some attenuation will take place 
at the lossy resonator. For backward propagation, light will first couple to the lossy resonator where it will experience attenuation. When the 
light reaches the gain microdisc, the accumulated energy there is not enough to appreciably red-shift the resonance. Thus light is 
transmitted to the left port through the resonance and experience a transmission dip at $\lambda_{\chi=0}^{\cal PT}$. Therefore a non-
reciprocal light transport at the wavelength $\lambda_{\chi=0}^{\cal PT}$ is observed. The asymmetric transport is further amplified due 
to the fact that the Fano resonance lineshape for a left and right incident waves can be different. 

\begin{figure}
   \includegraphics[width=1\linewidth, angle=0]{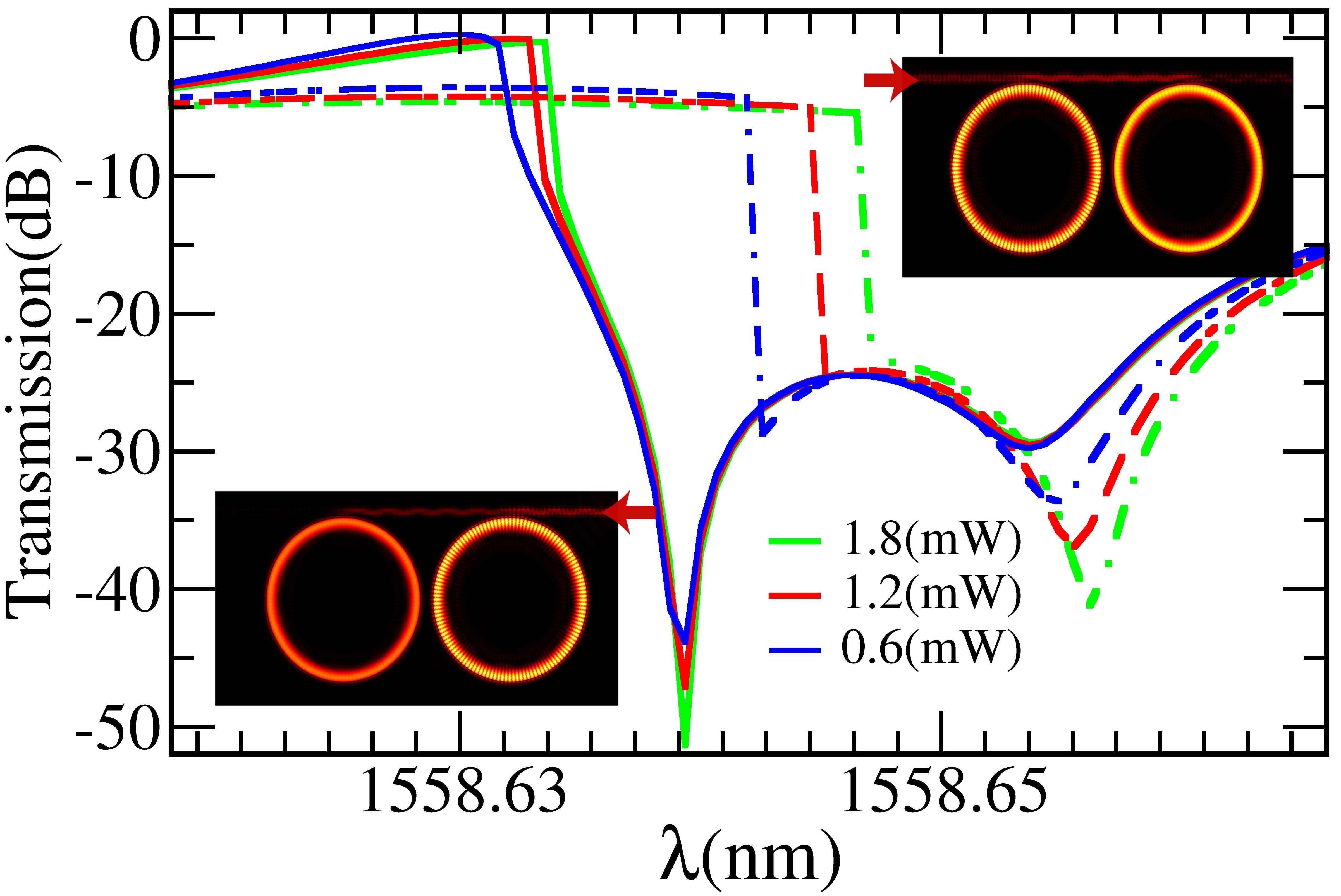}
    \caption{(Color online) Transmittance for different input powers. The asymmetric transport is maintained for a broad range of input 
power levels. Dashed lines correspond to $T_L(\lambda)$ (gain side) while solid lines to $T_R(\lambda)$ (lossy side). The transport 
asymmetry is as high as 46.5 dB without compromising the outgoing optical intensity which is as high as -5 dBs.}
\label{fig:fig3}
\end{figure}

In Fig. \ref{fig:fig3} we further analyze the dependence of the transmittance on the level of the input power. We find that the circuit is stable 
to variations of the input power; a feature that is desirable from the engineering perspective. For these simulations we assume an imaginary 
permittivity index $\epsilon'' = 0.00073$ i.e. slightly larger than the one used in Fig. \ref{fig:fig1}.This allow us to enhance further the transport 
asymmetry to values as high as 46.5 dBs. At the same time the outgoing signal $T_L(\lambda=\lambda_{\chi=0}^{\cal PT})$ is further 
amplified ($\approx -5$ dBs) with respect to the one found in Fig. \ref{fig:fig1} ($\approx -8$ dBs). This has to be contrasted with passive protocols 
where an increase figure of merit for isolation might lead to weaker outgoing signal \cite{LC11,FWVSNXWQ12}.

The asymmetric transport generated by the interplay of the nonlinear Fano resonances with ${\cal PT}$-symmetric elements calls for a simple 
theoretical understanding. The following heuristic model, similar in spirit to the so-called Fano-Anderson model \cite{M93,MFK10} that is used 
to describe the creation of (non-linear) Fano resonances, provides some quantitative understanding of the COMSOL simulations shown in Figs. 
\ref{fig:fig1},\ref{fig:fig3}. Our model is described by the following sets of differential equations:
\begin{eqnarray}
\label{dif1}
i\dot{\phi_{n}} &=&-\{C(\phi_{n-1} + \phi_{n+1}) + V_{G} \phi_G \delta_{n,0} + V_{L} \phi_L \delta_{n,N} \}\nonumber\\
i\dot{\phi_G} &=& -\{ (E - i\gamma) \phi_G + \chi |\phi_G|^{2} \phi_G + V_{G} \phi_{0} \}\nonumber\\
i\dot{\phi_L} &=&-\{ (E + i\gamma) \phi_L + \chi |\phi_L|^{2} \phi_L + V_{L} \phi_{N}\}
\end{eqnarray}
Equations (\ref{dif1}) describe the interaction of two subsystems. The first one is a linear chain of couple sites with coupling constant $C$ and
on-site complex field amplitudes $\phi_n$. This system supports propagating plane waves with dispersion $\omega(k)=2C cos q$. The second 
subsystem consists of two defect states $\phi_G$ (gain) and $\phi_L$ (loss) with on-site energy $E\mp i\gamma$ respectively. The two subsystems
interact with one another at the sites $n=0,N$ via the coupling coefficients $V_{G/L}$.

We assume elastic scattering processes for which the stationary solutions take the form $\phi_{n} = A_{n} e^{i \omega t};\,\, \phi_G = A_g e^{i \omega t};
\,\,\phi_L = A_L e^{i \omega t}$. Substitution in Eqs. (\ref{dif1}) leads to
\begin{eqnarray}
\label{dif2}
\omega A_{n} &=& C(A_{n-1} + A_{n+1}) + V_{G} A_G \delta_{n,0} + V_{L} A_L \delta_{n,N}\nonumber\\
\omega A_G &=& E A_G - i\gamma A_G + \chi |A_G|^{2} A_G + V_{G} A_{0} \nonumber\\
\omega A_L &=& E A_L + i\gamma A_L + \chi |A_L|^{2} A_L + V_{L} A_{N}
\end{eqnarray}
We consider a left incident wave. In this case we have
\begin{equation}
\label{ans}
A_{n} = \left\{
\begin{array}{cc}
I e^{i q n} + r e^{-i q n} & n \leq 0\\
\alpha e^{i q n} + \beta e^{-i q n} & \ 0 \leq n \leq N\\
t e^{i q n} & N \leq n
\end{array}\right .
\end{equation}
where $I,r,t$ represent the incident, reflected and transmitted wave amplitudes far from the defect sites. Substituting the 
above scattering conditions in Eqs. (\ref{dif2}) and using the continuity conditions at the defect sites $n=0,N$, we get after some 
straightforward algebra
\begin{equation}
\label{tr_r}
r_L  = i\frac{V_{G} A_G + V_{L} A_Le^{iqN} }{2 C \sin{q}};
t_L  = (I +i \frac{V_{G} A_G + V_{L} A_L e^{-iqN}}{2 C \sin{q}})
\end{equation}
The unknown amplitudes $A_G,A_L$ can be found in terms of the input amplitute $I$ by utilizing Eqs. (\ref{dif2}b,c). Specifically we 
get the following set of nonlinear equations
\begin{eqnarray}
\label{eq12}
(-\omega  + E  - i\gamma) A_G + \chi |A_G|^{2} A_G + V_{G} (I + r_L)  &=& 0\quad\\
(-\omega + E + i\gamma) A_L + \chi |A_L|^{2} A_L + V_{L} e^{i q N} t_L& =& 0\nonumber
\end{eqnarray}
which can be solved numerically, after substituting $r_L$ and $t_L$ from Eqs. (\ref{tr_r}).

\begin{figure}
   \includegraphics[width=1\linewidth, angle=0]{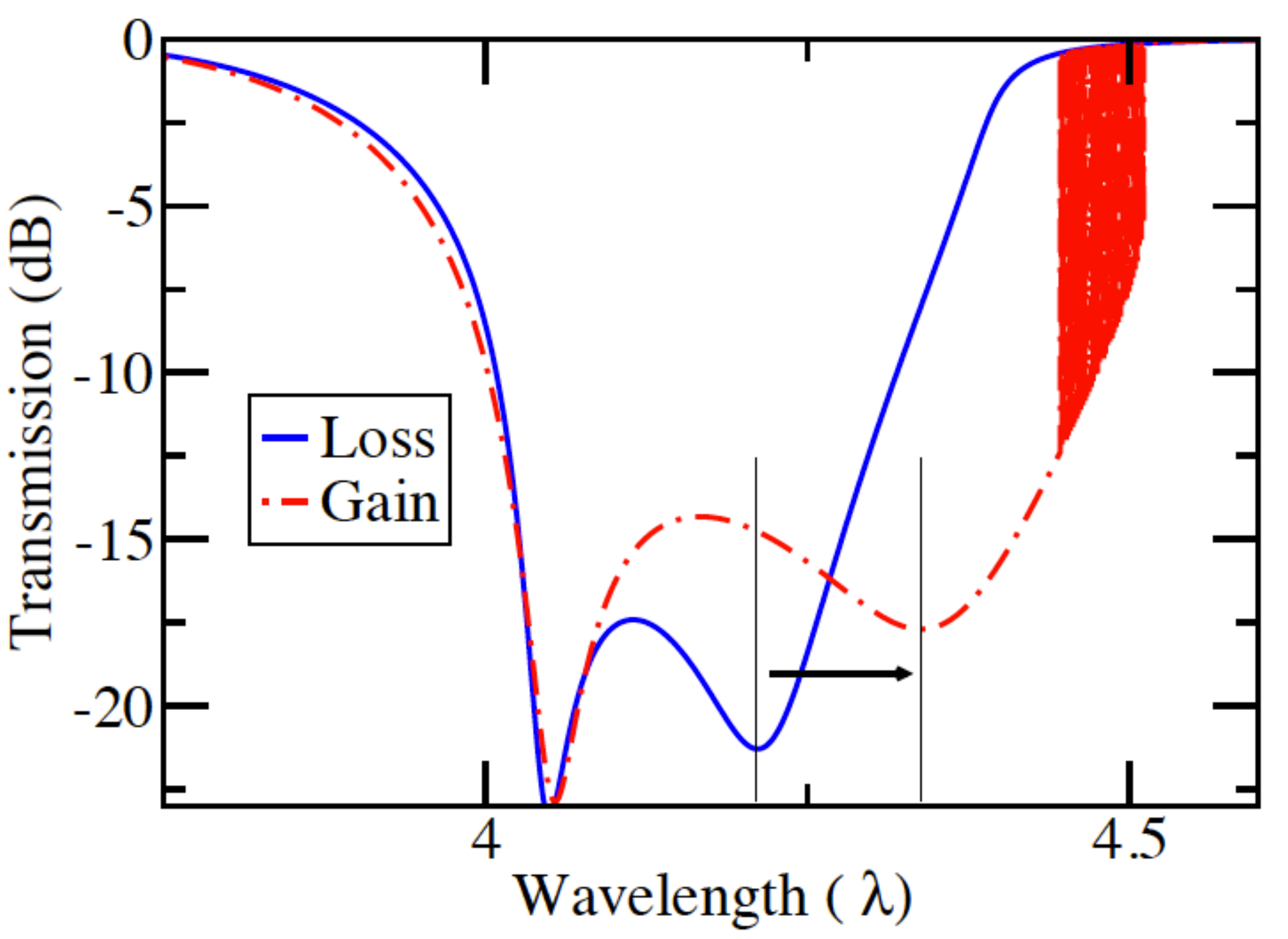}
    \caption{(Color online) Left (gain) $T_L$ and right (loss) $T_R$ transmittances for the theoretical model of Eq. (\ref{dif1}). Notice the red-shift
associated with $T_L$ (pointed with a black arrow), in the neighborhood of the second Fano resonance. In this domain the $T_L> T_R$. The parameters
used in this simulation are $V_G=V_L=0.5$, $N=1$, $\chi=0.0125$ and $\gamma=0.02$. The red-shadowed area on the right of the graph 
around $\lambda\approx 4.5$, corresponds to a bi-stability behavior which however is away from the Fano resonance regime and thus does not 
affect asymmetric transport.}
\label{fig:fig4}
\end{figure}

The transmittance and reflectance for a left incident wave is defined as $T_L=|t_L/I|^2$ and $R_L=|r_L/I|^2$ respectively. In a similar 
manner one can also define the transmittance $T_R=|t_R/I|^2$ and reflectance $R_R=|r_R/I|^2$ for a right incident wave. The associated 
$t_R, r_R$ are given by the same expressions as Eq. (\ref{tr_r}, \ref{eq12}) with the substitution of $\gamma\rightarrow -\gamma; 
V_G\rightarrow V_L$ and  $V_L\rightarrow V_G$.

An analysis of the structure of Eq. (\ref{tr_r}) can explain the origin of Fano resonances. Specifically, 
we note that the transmission amplitude in Eq. (\ref{tr_r}) consists of two terms: the first one is associated with a scattering process 
associated with a propagating wave that directly passes through the chain without coupling to any of the defect states. The second term
describes an indirect path for which the wave will first visit the two defects, thus exciting the Fano states, return back, and continue 
with the propagation. These two paths are the ingredients of the Fano resonances observed in Figs. \ref{fig:fig1},\ref{fig:fig3}.

In Fig. \ref{fig:fig4} we report a representative set of transmission curves for a left/right incident wave for the model of Eq. (\ref{dif1}).
The model capures the qualitative features and origin of the asymmetric transport observed in the case of the photonic circuit of Fig. \ref{fig:fig1}.
Specifically, we find that both the shape and the position of the Fano resonances depend on the direction of the incident wave. Moreover 
for a left (gain-side) incoming wave, a red-shift in the transmittance resonances is found  (see the neigborhood of the second Fano 
resonance in Fig. \ref{fig:fig4}) which leads to an asymmetric transport.

{\it Conclusions -} In conclusion, we have introduce a new type of ${\cal PT}$-symmetric Fano resonances with a line-shape and a resonance
position that depends on the direction of the incident wave. The photonic circuit that allows for such resonances consists of two ${\cal PT}$-
symmetric microdiscs side-coupled to a waveguide. The proposed configuration guarantee not only high asymmetry but also a significant level 
of transmittance. Our proposal utilizes existing materials already used in optical integrated circuitry processing and does not require magnetic 
fields, or other external elements like polarizers. The efficiency of the asymmetric transport for a broad input power range and low values of input power 
within which our device performs may be sufficient for on-chip photonic applications. A problem with the simple design of Fig. \ref{fig:fig1} is that 
the asymmetric transport only occurs in the vicinity of the Fano resonances (narrow band). This problem can be addressed by using more 
sophisticated photonic structures, for instance, those involving more than one ${\cal PT}$-symmetric dimer-like resonators side coupled to the
waveguide bus.



\end{document}